\documentclass[twocolumn,aps,superscriptaddress,showpacs]{revtex4}
\usepackage{amssymb}
\usepackage{amsmath}
\usepackage{graphicx}
\usepackage[normalem]{ulem}
\usepackage[dvips]{color}
\usepackage{multirow}
\usepackage{appendix}

\makeatletter

\newcommand{\Rmnum}[1]{\expandafter\@slowromancap\romannumeral #1@}
\makeatother

\begin{document}

\title{Effects of in-medium nucleon-nucleon cross section on collective flow and nuclear stopping in heavy-ion collisions in the Fermi-energy domain}
\author{Pengcheng Li}
\affiliation{School of Science, Huzhou University, Huzhou 313000, China}
\affiliation{School of Nuclear Science and Technology, Lanzhou University, Lanzhou 730000, China}
\author{Yongjia Wang \footnote{corresponding author: wangyongjia@zjhu.edu.cn}}
\affiliation{School of Science, Huzhou University, Huzhou 313000, China}
\author{Qingfeng Li \footnote{corresponding author: liqf@zjhu.edu.cn}}
\affiliation{School of Science, Huzhou University, Huzhou 313000, China}
\affiliation{Institute of Modern Physics, Chinese Academy of Sciences, Lanzhou 730000, China}
\author{Chenchen Guo}
\affiliation{Sino-French Institute of Nuclear Engineering and Technology, Sun Yat-sen University, Zhuhai 519082, China}
\author{Hongfei Zhang}
\affiliation{School of Nuclear Science and Technology, Lanzhou University, Lanzhou 730000, China}
\date{\today}

\begin{abstract}
With the newly updated version of the ultrarelativistic quantum molecular dynamics (UrQMD) model, a systematic investigation of the effects of in-medium nucleon-nucleon ($NN$) elastic cross section on the collective flow and the stopping observables in $^{197}\text{Au}+^{197}\text{Au}$ collisions at beam energies from 40 to 150 MeV/nucleon is performed. Simulations with the medium correction factor $\mathcal{F}=\sigma^{\text{in-medium}}_{NN}/\sigma^{\text{free}}_{NN}=0.2,~0.3,~0.5$, and the one obtained with the FU3FP1 parametrization which depends on both the density and the momentum are compared to the FOPI and INDRA experimental data. It is found that, to best fit the experimental data of the slope of the directed flow and the elliptic flow at mid-rapidity as well as the nuclear stopping, the correction factor $\mathcal{F}$=0.2 and 0.5 are required for reactions at beam energies of 40 and 150 MeV/nucleon, respectively. While calculations with the FU3FP1 parametrization can simultaneously reproduce these experimental data reasonably well. And, the observed increasing nuclear stopping with increasing beam energy in experimental data can also be reproduced by using the FU3FP1 parametrization, while the calculated stopping power in Au+Au
collisions with beam energies from 40 to 150 MeV$/$nucleon almost keeps constant when take $\mathcal{F}$ equal to a fixed value.

\end{abstract}


\maketitle

\section{Introduction}
\label{introduction}

Investigations of the equation of state (EoS) of nuclear matter and the nucleon-nucleon ($NN$) cross section have drawn much attention during the past several decades in both nuclear physics and astrophysics due that they are essential for understanding many phenomena in nuclear structures and reactions, as well as in astrophysical nuclear processes \cite{Baran:2004ih,Steiner:2004fi,Lattimer:2006xb,DiToro:2010ku,Li:2014oda}. So far, the stiffness of the EoS for isospin symmetric nuclear matter has been relatively well understood, though there still remains some uncertainties for further improvement \cite{Khan:2013mga,Stone:2014wza,Fevre:2015fza,yjw-plb}. Concerning the $NN$ cross section, in free space, its information has been well measured by experiments, but in the nuclear medium, it should be relied on comparison of the theoretical calculations to experimental data of heavy ion collisions (HICs). It is well known that the in-medium $NN$ cross section is suppressed when compared to the free one, however, the degree of this suppression is still far from being completely pinned down.

Theoretically, the in-medium $NN$ cross section can be calculated by using different methods, e.g., the Dirac-Brueckner approach with the Bonn potential \cite{lgq14}, the Dirac-Brueckner-Hartree-Fock approach with realistic nucleon-nucleon potential \cite{Sammarruca:2005tk}, the relativistic Brueckner-Hartree-Fock model \cite{HFZ07,CF,HJS}, the T-matrix approach \cite{Alm:1995chb,WGL}, and the relativistic BUU (RBUU) microscopic transport theory with the effective Lagrangian \cite{Mao:1994zza,Li:2003vd}. Alternatively, the detailed information of the in-medium $NN$ cross section can be deduced from the comparison of observables of HICs with corresponding transport model simulations. For instance, by studying the balance energy, the collective flow and the stopping power with microscopic transport models, strong evidence for the reduction of $NN$ cross section in nuclear medium have been confirmed in HICs at intermediate energies \cite{Xu:1992zza,gd10,Li:1999bh,Magestro:2000ba,ljy12,dp,tg,Li:2005iba,qfli,Zhang:2007gd,yuany2010,dds,Zhang:2011yv,Feng:2011eu,YZ10,Kaur:2016eaf,js24,Zbiri:2006ts}. The frequently used transport models for the HICs at low and intermediate energies are the Quantum Molecular Dynamics (QMD) model \cite{Aichelin:1991xy} and the Boltzmann- (Vlasov) Uehling-Uhlenbeck (BUU, VUU) model \cite{GF7}. Usually, parameterized in-medium $NN$ elastic cross sections are adopted in transport model for simplicity. For example, $\sigma_{NN}^{\text{in-medium}}=(1-\eta\rho/\rho_{0})\sigma_{NN}^{\text{free}}$ with $\eta=0.2$ has been used in models and it was found to better reproduce the flow and stopping experimental data \cite{gd10,Zhang:2007gd}.
In the pBUU model, the in-medium $NN$ cross section is implemented in the form $\sigma_{NN}^{\text{in-medium}}=0.85\rho^{-2/3}/\tanh(\frac{\sigma^{\text{free}}}{0.85\rho^{-2/3}})$ \cite{dds}. While in the isospin-dependent Boltzmann-Uehling-Uhlenbeck (IBUU) model and the Lanzhou Quantum Molecular Dynamics (LQMD) model, the in-medium $NN$ cross section is reduced by a factor $\mathcal{F}=\sigma_{NN}^{\text{in-medium}}/\sigma_{NN}^{\text{free}}=(\mu_{NN}^{*}/\mu_{NN})^2$, where $\mu_{NN}^{*}$ and $\mu_{NN}$ are the $k$-masses of the colliding nucleon partners in the nuclear medium and in free space \cite{Li:2005iba,Feng:2011eu}, respectively. In Ref.~\cite{Zbiri:2006ts}, the ratio between the mean squared rapidity variances in the impact parameter direction and in the longitudinal direction $R=\frac{<y^2_x>}{<y^2_z>}$ as a function of the charge number of the fragments produced in Au+Au collisions at 150 MeV$/$nucleon was investigated, the global trend with increasing charge number can be reproduced by the ``standard" QMD model (of J. Aichelin \emph{et al}.) in which the free $NN$ cross section is considered. However, the calculated $R$ of Z=1 fragments is much larger than the measured data, which also implies that a reduction in the $NN$ cross section might be necessary. Hence, the $NN$ cross section in different transport models behaves differently and deserves further investigation.

Recently, a systematic study of nuclear stopping of protons from central HICs at Fermi-energy domain was performed by the INDRA collaboration using the powerful INDRA 4$\pi$ array, it was demonstrated that the mean free path $\lambda_{NN}$=9.5$\pm$2 fm at the beam energy of 40 MeV$/$nucleon and $\lambda_{NN}$=4.5$\pm$1 fm at 100 MeV$/$nucleon, based on the assumption that $\sigma_{NN}\approx 1/(\rho \lambda_{NN})$, the correction factor $\mathcal{F}=\sigma_{NN}^{\text{in-medium}}/\sigma_{NN}^{\text{free}}$ is then deduced to be $0.16\pm0.04$ at 40 MeV$/$nucleon and  $0.5\pm0.06$ at 100 MeV$/$nucleon, respectively \cite{OL13}. These stopping data provide a new  opportunity to re-visit the in-medium $NN$ cross section by using transport models \cite{MYG2017,BZ2016,XYZ2017,BKK2015}. In our previous works, within the ultrarelativistic quantum molecular dynamics (UrQMD) model, it is found that the collective flow of light clusters and the nuclear stopping can be reproduced reasonable well with the consideration of a medium correction factor $\mathcal{F}$ (which depends on density and momentum) on the free $NN$ cross section \cite{lq1119,wyj1417,Li:2005gfa}. Thus, it is of great interest to know the difference between the $\mathcal{F}$ extracted from experimental data \cite{OL13} and the one adopted currently in the UrQMD model.

In this work, with the updated potential version of the UrQMD model, we investigate the influence of the in-medium $NN$ cross section on collective flow and stopping of protons from Au+Au collisions at Fermi-energy domain by considering various corrections. In the next section the UrQMD model, the in-medium correction factors of $NN$ elastic cross section, as well as flow and stopping quantities are introduced briefly. In Sec. III, effects on both observables of free protons and hydrogen isotopes are shown and discussed. Finally, a summary is given in Sec. IV.

\section{Model description and observables}
\label{model}

In the UrQMD model, each nucleon is represented by Gaussian wave packet with the width parameter $L$ in phase space~\cite{Bass:1998ca}. Usually, $L=2$ fm$^2$ is chosen for simulating Au+Au collisions. The centroids of coordinate $\textbf{r}_i$ and momentum $\textbf{p}_i$ of nucleon $i$ are propagated according to
\begin{eqnarray}
\dot{\textbf{r}}_{i}=\frac{\partial  \langle H  \rangle}{\partial\textbf{ p}_{i}},
\dot{\textbf{p}}_{i}=-\frac{\partial  \langle H \rangle}{\partial \textbf{r}_{i}}.
\end{eqnarray}
Here, {\it $\langle H \rangle$} is the total Hamiltonian function of the system, it comprises the kinetic energy $T$ and the effective interaction potential energy $U$. For studying HICs at intermediate energies, the following density and momentum dependent potential was frequently used in QMD-like models \cite{Aichelin:1991xy,Hartnack:1997ez,Li:2005gfa},
\begin{equation}\label{eq2}
U=\alpha \cdot (\frac{\rho}{\rho_0})+\beta \cdot (\frac{\rho}{\rho_0})^{\gamma} + t_{md} \ln^2[1+a_{md}(\textbf{p}_{i}-\textbf{p}_{j})^2]\frac{\rho}{\rho_0}.
\end{equation}
Here $\alpha$=-393 MeV, $\beta$=320 MeV, $\gamma$=1.14, $t_{md}$=1.57 MeV, and $a_{md}=0.0005$ MeV$^{-2}$ are chosen, which yields the incompressibility $K_0$=200 MeV for isospin symmetric nuclear matter. In recent years, to better describe the recent experimental data at intermediate energies and following present progress on determining the nuclear symmetry energy, the surface, and surface asymmetry energy term, as well as the bulk symmetry energy term obtained from the Skyrme potential energy density functional have been further introduced to the present version~\cite{wyj1417,Wang:2014aba}, which reads as

\begin{equation}\label{urho}
\begin{aligned}
u_{Skyrme}=&u_{sur}+u_{sur,iso}+u_{sym}\\
&=\frac{g_{\text{sur}}}{2\rho_{0}}(\nabla\rho)^{2}+\frac{g_{\text{sur,iso}}}{2\rho_{0}}[\nabla(\rho_{n}-\rho_{p})]^{2}\\
&+(A_{\text{sym}}\frac{\rho^{2}}{\rho_{0}}+B_{\text{sym}}\frac{\rho^{\eta+1}}{\rho_{0}^{\eta}}+C_{\text{sym}}\frac{\rho^{8/3}}{\rho_{0}^{5/3}})\delta^2.
\end{aligned}
\end{equation}
Here, $\delta=(\rho_{n}-\rho_{p})/(\rho_{n}+\rho_{p})$ is the isospin asymmetry defined by the neutron ($\rho_n$) and proton ($\rho_p$) densities. And, the parameters $g_{\text{sur}}$, $g_{\text{sur,iso}}$, $A_{sym}$, $B_{sym}$, and $C_{sym}$ are related to the Skyrme parameters via
\begin{eqnarray}
  \frac{g_{sur}}{2} &=& \frac{1}{64}(9t_{1}-5t_{2}-4x_{2}t_{2})\rho_{0}, \\
  \frac{g_{sur,iso}}{2} &=& -\frac{1}{64}[3t_{1}(2x_{1}+1)+t_{2}(2x_{2}+1)]\rho_{0},\\
  A_{sym} &=& -\frac{t_{0}}{4}(x_{0}+1/2)\rho_{0}, \\
  B_{sym} &=& -\frac{t_{3}}{24}(x_{3}+1/2)\rho_{0}^{\eta}, \\
  C_{sym} &=& \frac{1}{24}(\frac{3\pi^{2}}{2})^{2/3}\rho_{0}^{5/3}\Theta_{sym},
\end{eqnarray}
where $\Theta_{sym}=3t_{1}x_{1}-t_{2}(4+5x_{2})$ \cite{wyj1417}. In this work, the SV-sym34 force, in which $g_{\text{sur}}=18.2$ MeV fm$^{2}$, $g_{\text{sur,iso}}=8.9$ MeV fm$^{2}$, $A_{sym}=20.3$ MeV,
$B_{sym}=14.4$ MeV, and
$C_{sym}=-9.2$ MeV, and the slope parameter of the symmetry energy $L=80.95$ MeV, is employed.

The directed $v_{1}$ and elliptic $v_{2}$ flows are the two of most frequently studied observables in HICs, which can be deduced from the Fourier expansion of the azimuthal distribution of detected particles \cite{WR23}, and reads as,
\begin{equation}\label{v1}
  v_{1}\equiv \langle cos(\phi)\rangle=\left\langle\frac{p_{x}}{p_{t}}\right\rangle,
\end{equation}
\begin{equation}\label{v2}
  v_{2}\equiv \langle cos(2\phi)\rangle=\left\langle\frac{p_{x}^{2}-p_{y}^{2}}{p_{t}^{2}}\right\rangle,
\end{equation}
in which $p_{x}$ and $p_{y}$ are the two
components of the transverse momentum $p_{t}=\sqrt{p_{x}^{2}+p_{y}^{2}}$. And the angle brackets in Eq.\ref{v1} and Eq.\ref{v2} indicate an average over all considered particles from all events.
Besides the directed and elliptic flows, the nuclear stopping power which characterizes the transparency of the colliding nuclei is another important observable and can be defined with different quantities.
A possible measurement of the degree of stopping is $vartl$ (proposed by the FOPI collaboration \cite{WR23}) which is defined as the ratio of the variances of the transverse to those of the longitudinal rapidity distribution, reads as,
\begin{equation}
  vartl=\frac{\langle y_{x}^{2}\rangle}{\langle y_{z}^{2}\rangle},
\end{equation}
here
\begin{equation}
\langle y_{x,z}^{2}\rangle=\frac{\sum(y_{x,z}^{2}N_{y_{x,z}})}{\sum N_{y_{x,z}}},
\end{equation}
where $\langle y_{x}^{2}\rangle$ and $\langle y_{z}^{2}\rangle$ are the variances of the rapidity distributions of particles in the $x$ and $z$ directions, respectively. Another quantity $R_{E}$ is also used to measure the stopping power, which was proposed by the INDRA collaboration, and defined as the ratio of transverse to parallel energy, reads as,
\begin{equation}
  R_{E}=\frac{\sum E_{\bot}}{2\sum E_{\|}},
\end{equation}
where $E_{\bot}$ ($E_{\|}$) is the transverse (parallel) kinetic energy of particles in center-of-mass system \cite{GL4}. Apparently, one can expect that for full stopping, both $R_{E}$ and $vartl$ values will to be unity, while they will be zero for full transparency.

The in-medium $NN$ elastic cross section is treated to be factorized as the product of a medium correction factor $\mathcal{F}(\rho, p)$ and the free cross section and reads,
\begin{equation}
\sigma_{NN}^{\text{in-medium}}=\mathcal{F}(\rho,p)*\sigma_{NN}^{\text{free}}
\end{equation}
with
\begin{equation}
\mathcal{F}(\rho,p)=\frac{\lambda+(1-\lambda)e^{-\rho/\rho_0/\zeta}-f_{0}}{1+(p_{NN}/p_{0})^{\kappa}}+f_{0}.
\end{equation}
Where $p_{NN}$ is the relative momentum of two colliding nucleons. In this work, the $\lambda=1/3$, $\zeta=1/3$, $f_{0}=1$, $p_{0}=0.425$ GeV/$c$, and $\kappa=5$, which corresponds to the FU3FP1 parametrization used in Ref. \cite{lq1119}. In addition, if $p_{NN}$ is larger than 1 GeV/$c$, $\mathcal{F}(\rho,p)$ is set to be unity.
\begin{figure}[htbp]\centering
\includegraphics[width=0.48\textwidth]{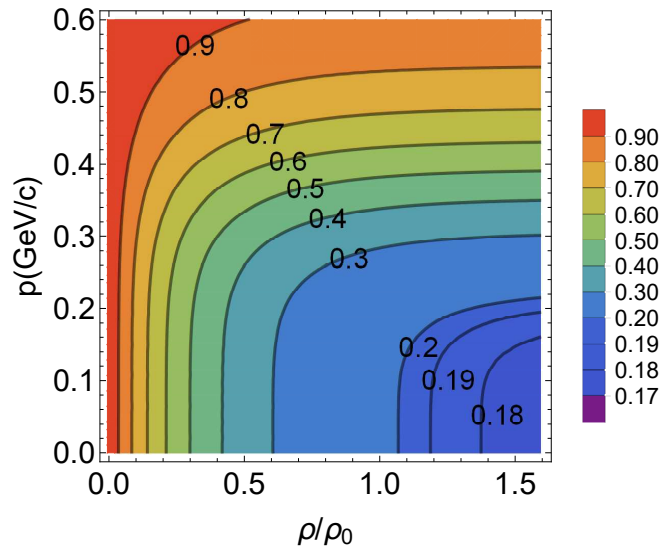}
\caption{(Color online) The in-medium correction coefficient $\mathcal{F}$ obtained from the FU3FP1 parametrization as functions of density and momentum.}
\label{fig1}
\end{figure}

The in-medium correction factor $\mathcal{F}(\rho,p)$ obtained
from the FU3FP1 parametrization is displayed in Fig.\ref{fig1}, as functions of both reduced density $\rho/\rho_{0}$ and momentum $p$ (here $p=p_{NN}$). It is seen that values of $\mathcal{F}$ obtained within (0.3$\le$$\rho/\rho_0$$\le$1.5, 0$\le$$p$$\le$0.4 GeV/$c$) cover the results obtained from Ref. \cite{OL13}.
In order to evaluate the effect of $\mathcal{F}$ values from the FU3FP1 parametrization and suggested in Ref. \cite{OL13} on flow parameters and nuclear stopping discussed above, three fixed in-medium correction factors $\mathcal{F}=0.2,~0.3$, and 0.5 are further considered for this work.

\section{Results and discussions}
\label{results}
 In order to have enough statistics for the analysis of the result, more than 450 thousand Au+Au events within the impact parameter $b=0-7.5$ fm at each beam energy (40, 50, 60, 80, 90, 100, 120, or 150 MeV$/$nucleon) are simulated. An isospin-dependent minimum span tree algorithm (iso-MST) is used to recognize fragments. Nucleons with relative distances smaller than $R_{0}$ and relative momenta smaller than $P_{0}$ are considered to belong to the same cluster. With proper set of these parameters, fragment mass distribution in intermediate energies HICs can be reproduced \cite{hk31,scpma58,pp2011}. In the present work, $R_{0}$ and $P_{0}$ are set to $R_0^{pp}=2.8$ fm, $R_0^{nn}=R_0^{np}=3.8$ fm and $P_0 =0.25$ GeV/$c$. Although the (iso-)MST method has been widely used in transport model to recognize fragments, the values of these coalescence model parameters ($R_0$ and $P_0$) are different in different models, see e.g., Refs.~\cite{Aichelin:1991xy,Zhang:2011yv,scpma58,pp2011,Zhang:2012qm,WR5}. Meaningful constraint on the in-medium $NN$ cross section can be extracted from transport calculations only if these parameters which are not under full control do not apparently affect the observable of interest. It was found that the influence of the coalescence model parameters on the collective flow of free protons is relatively weak, see e.g., Refs.~\cite{wyj1417,Kumar:2013xbw}. In Ref. \cite{Zhang:2012qm}, the influence of coalescence model parameters on the degree of nuclear stopping was studied, with a maximum set of these parameters, the $vartl$ for Z=1-6 particles obtained with iso-MST is about 7\% larger than that with the isospin-independent MST. Thus in this work, only the collective flow and nuclear stopping of free protons are used to extract the in-medium $NN$ cross section.

\subsection{Collective flow}
\begin{figure}[htbp]\centering
\includegraphics[width=0.5\textwidth]{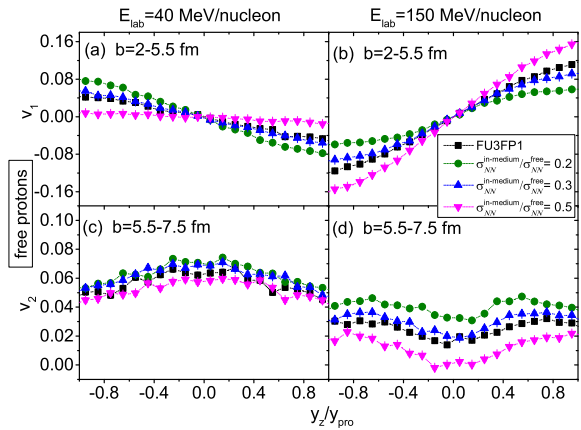}
\caption{(Color online) The directed flow $v_{1}$ and elliptic flow $v_{2}$ of free protons from Au+Au collisions at $E_{\text{\text{lab}}}$ = 40 (left panels) and 150 MeV/nucleon (right panels), as a function of the reduced rapidity $y_{z}/y_{\text{pro}}$. The impact parameters are chosen to be b=2-5.5 fm and b=5.5-7.5 fm for the directed and elliptic flows, respectively. The calculations performed with the in-medium correction factors $\mathcal{F}=0.2$ (circles), $\mathcal{F}=0.3$ (up-triangles), and $\mathcal{F}=0.5$ (down-triangles) are compared with calculations using the FU3FP1 parametrization on the in-medium $NN$ cross section (squares).}
\label{fig2}
\end{figure}
Fig.\ref{fig2} shows the directed ($v_{1}$) and elliptic ($v_{2}$) flows of free protons as a function of the reduced longitudinal rapidity ($y_{z}/y_{pro}$) from $^{197}$Au+$^{197}$Au collisions at 40 MeV/nucleon and 150 MeV/nucleon with different medium correction factors ($\mathcal{F}$). It can be clearly seen that both $v_{1}$ and $v_{2}$ are affected by the medium correction factor $\mathcal{F}$. The difference in $v_1$ or $v_2$ with different $\mathcal{F}$ becomes more evident at $E_{\text{\text{lab}}}$ =150 MeV/nucleon than that at 40 MeV/nucleon, because at the higher beam energy one expects that the collision term plays the more important role. Further, the value of slope of $v_1$ at mid-rapidity ($y_{z}/y_{\text{pro}}$=0) increases and the value of $v_2$ at $y_{z}/y_{\text{pro}}$=0 decreases with increasing $\mathcal{F}$. This is due to the fact that the increasing collision number makes nucleons more likely undergo a bounce-off (positive $v_1$ slope) motion and squeeze-out (negative $v_2$) pattern. In addition, it is interesting to see that both the $v_1$ and $v_2$ obtained with the FU3FP1 are very close to that obtained with $\mathcal{F}=0.3$.

\begin{figure}[htbp]\centering
\includegraphics[width=0.48\textwidth]{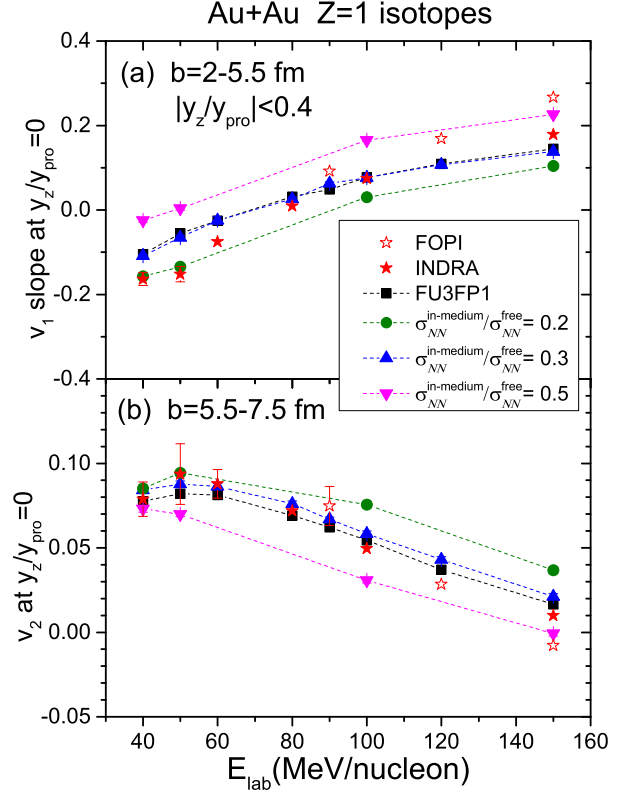}
\caption{(Color online) Beam energy dependence of $v_{1}$ slope [panel (a)] and $v_{2}$ [panel (b)] at mid-rapidity for hydrogen isotopes from $^{197}\text{Au}+^{197}\text{Au}$ collisions. The chosen impact parameters are b=2-5.5 fm and b=5.5-7.5 fm for $v_1$ and $v_2$, respectively. The $v_1$ slope is obtained with assuming $v_{1}(y_{0})=v_{11}y_{0}+v_{13}y_{0}^{3}+c$ in the range of $|y_{0}|=|y_{z}/y_{\text{pro}}|<0.4$. Calculated results with four in-medium correction factors are presented by different lines as indicated, the FOPI experimental data (open stars) and the INDRA experimental data (solid stars) are taken from Ref. \cite{AA25}.}
\label{fig3}
\end{figure}

To quantitatively estimate the influence of the medium correction factor on the directed flow and elliptic flow, the $v_1$ slope value and the $v_2$ value at mid-rapidity for hydrogen isotopes calculated with different $\mathcal{F}$ values are compared to the FOPI and INDRA experimental data taken from Ref. \cite{AA25}, and shown in Fig. \ref{fig3}. Similar to the results shown in Fig. \ref{fig2}, the value of $v_1$ slope increases and the value of $v_2$ decreases with increasing $E_{lab}$. Once again, the results obtained with the FU3FP1 and with $\mathcal{F}=0.3$ overlap appreciably in the whole beam energy region. And, as a whole, the FU3FP1 and $\mathcal{F}=0.3$ cases best describe the experimental data among all calculations. It is also seen that, both the $v_1$ and $v_2$ observables can be reproduced well with $\mathcal{F}=0.2$ at $E_{\text{lab}}$= 40 MeV$/$nucleon. While, at $E_{\text{lab}}$= 150 MeV$/$nucleon, calculations with $\mathcal{F}=0.5$ lie between the FOPI and the INDRA experimental data. Therefore, our calculations on the $\mathcal{F}$ factor are quite similar to the results shown in Ref. \cite{OL13}.

\begin{figure}[htbp]\centering
\includegraphics[width=0.5\textwidth]{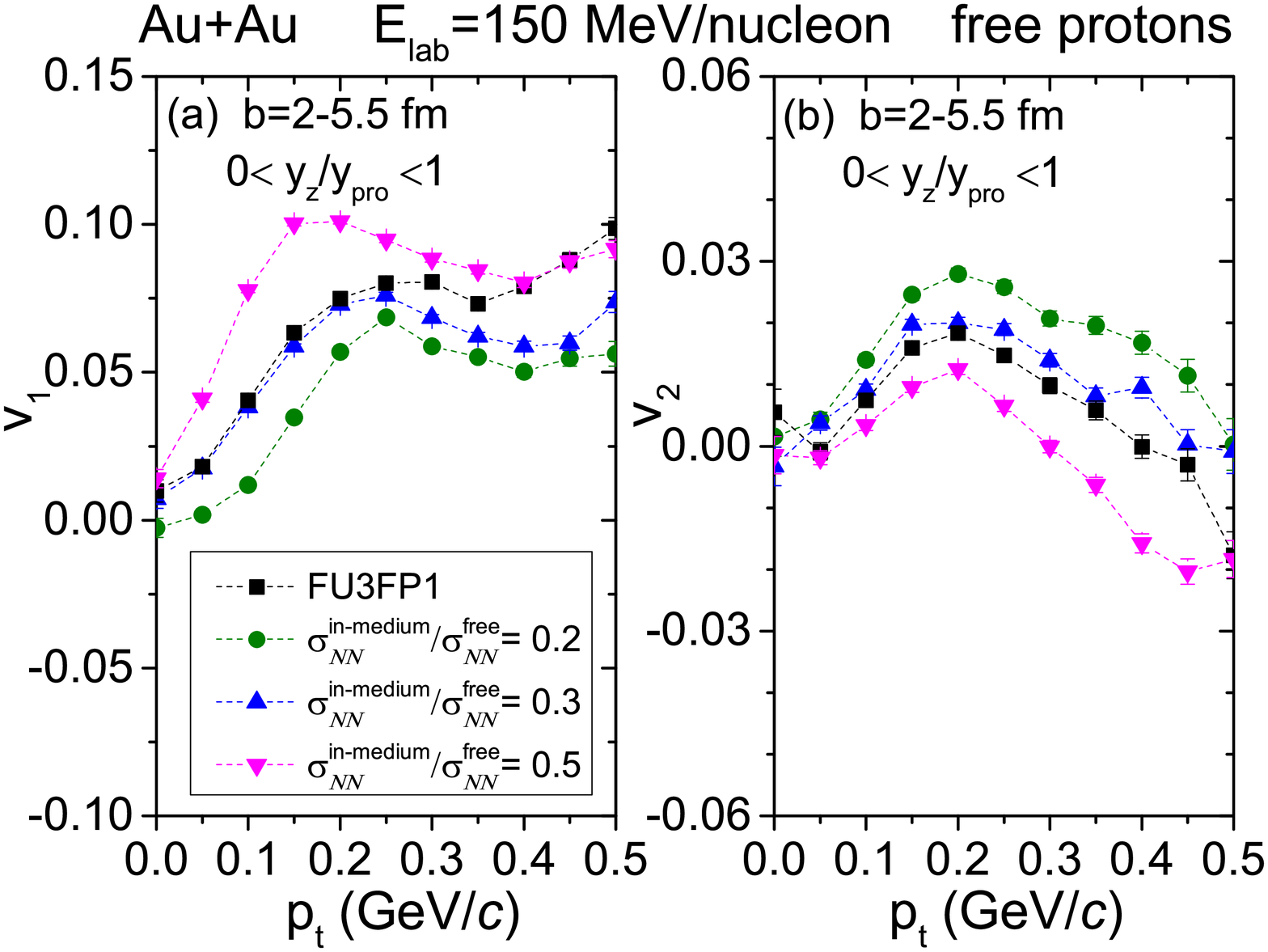}
\caption{(Color online) The directed flow $v_{1}$ (left) and elliptic flow $v_{2}$ (right) of free protons for $^{197}\text{Au}+^{197}\text{Au}$ collisions at the beam energy 150 MeV/nucleon, as a function of transverse momentum $p_{t}$. The impact parameter $b=2-5.5$ fm and the rapidity cut $0<y_z/y_{pro}<1$ are chosen for both $v_{1}$ and $v_{2}$.}
\label{fig4}
\end{figure}

To understand more clearly the difference caused by different medium correction factors on the collective flow, the transverse momentum $p_{t}$ dependence of the parameters $v_{1}$ and $v_{2}$ of free protons from $^{197}\text{Au}+^{197}\text{Au}$ collisions at $E_{lab}=150$ MeV/nucleon are exhibited in Fig. \ref{fig4}. First of all, with the increasing $\mathcal{F}$, the directed flow becomes larger and the elliptic flow becomes smaller as expected. Both the $v_1$ and $v_2$ obtained with the FU3FP1 parametrization and $\mathcal{F}=0.3$ are close to each other at low $p_t$, but the difference steadily increases with increasing $p_t$. It is known that particles with high $p_t$ usually emit early and experience only a fewer collisions and with a larger relative momentum. The medium correction factor obtained with the FU3FP1 parametrization maintains the momentum dependence, so that the medium suppression effect is weakened at high momenta. Thus the number of collision for the case of the FU3FP1 parametrization is larger than that for $\mathcal{F}=0.3$. It implies that the collective flow at high transverse momenta would be a promising probe for investigating the medium correction on the $NN$ cross section.

\subsection{Nuclear stopping}
\begin{figure}[htbp]\centering
\includegraphics[width=0.5\textwidth]{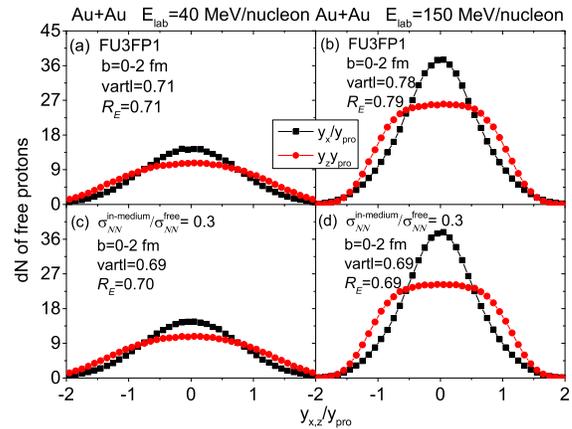}
\caption{(Color online) Yield distributions of free protons as functions of the reduced longitudinal ($y_{z}/y_{\text{pro}}$, circles) and transverse ($y_{x}/y_{\text{pro}}$, squares) rapidities from central (b = 0-2 fm) $^{197}\text{Au}+^{197}\text{Au}$ collisions at $E_{\text{lab}}$ = 40 (left panels) and 150 (right panels) MeV/nucleon. Calculations with the FU3FP1 parametrization [(a) and (b)] and $\mathcal{F}=0.3$ [(c) and (d)] are shown. The corresponding values for $vartl$ and $R_{E}$ are also indicated in each panel.}
\label{fig5}
\end{figure}

Besides the collective flow, the degree of nuclear stopping in HICs is another important observable which is also sensitive to the medium correction on the $NN$ cross section \cite{tg,lq1119,Zhang:2007gd,yuany2010}. In this work, we calculated the two quantities $R_{E}$ and $vartl$ from the same reaction. Fig. \ref{fig5} displays the yield distributions of free protons as functions of the reduced longitudinal and transverse rapidities for central $^{197}\text{Au}+^{197}\text{Au}$ collisions at $E_{\text{lab}}=40$ (left) and 150 MeV/nucleon (right). And, the FU3FP1 parametrization (top) and $\mathcal{F}=0.3$ (bottom) are selected for comparison. The extracted values of $vartl$ and $R_{E}$ are also given in each panel, and they are almost equal to each other at each beam energy as well as with both medium correction factors. We have checked that, although the values of $vartl$ and $R_E$ for free nucleons are almost equal to each other, the values of $R_E$ are usually smaller than $vartl$ for light fragments such as deuterons and tritons. Further, at $E_{\text{lab}}=40$ MeV/nucleon, the $vartl$ or $R_E$ obtained from the FU3FP1 parametrization are almost the same as that obtained with $\mathcal{F}=0.3$, while at $E_{\text{lab}}=150$ MeV/nucleon, the values of $vartl$ and $R_E$ obtained from the FU3FP1 are about 14\% larger that that from $\mathcal{F}=0.3$.

\begin{figure}[h]
\includegraphics[width=0.5\textwidth]{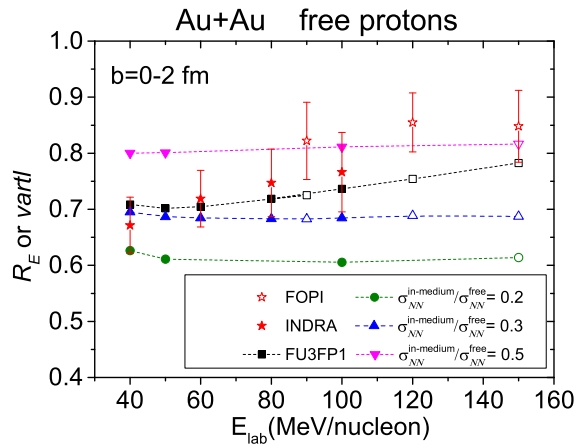}
\caption{(Color online) Beam energy dependence of the $R_{E}$ (solid symbols) and $vartl$ (open symbols) for free protons from central $^{197}\text{Au}+^{197}\text{Au}$ collisions. Calculations with four different medium correction factors (dashed lines with different symbols) are compared with the FOPI (open stars) and INDRA (solid stars) experimental data, which are taken from Ref. \cite{WR5} and Ref. \cite{OL13}, respectively.}
\label{fig6}
\end{figure}

Fig.\ref{fig6} displays the degree of nuclear stopping ($R_{E}$ or $vartl$) in central Au+Au collisions as a function of the beam energy. The results obtained with $\mathcal{F}=0.5$ are the largest (keep almost constant at $\sim0.8$) and those with $\mathcal{F}=0.2$ are the smallest ($\sim0.6$) of all. Again, this result from the nuclear stopping observables also consistently supports that the medium correction factors of about 0.2 and 0.5 are required for reasonably describing the degree of nuclear stopping in HICs at $E_{\text{lab}}$ = 40 and 150 MeV/nucleon, respectively. Further, the difference between the results from the FU3FP1 parametrization and from $\mathcal{F}=0.3$ (keep almost constant at $\sim0.7$) steadily increases with increasing beam energy. Calculations with FU3FP1 fit the experimental data quite well and reproduce the slightly increased stopping power with increasing beam energy, while others fail to reproduce the observed beam-energy dependence.

\section{Summary}
\label{summary}

Within the UrQMD model, the effects of the medium correction on the $NN$ elastic cross section on the collective flow and nuclear stopping in Au+Au at beam energies of 40-150 MeV$/$nucleon are investigated. Calculations performed with the medium correction factor $\mathcal{F}=\sigma_{NN}^{\text{in-medium}}/\sigma_{NN}^{\text{free}}$ of 0.2, 0.3, 0.5, and with the density- and momentum- dependent factor obtained from the FU3FP1 parametrization are compared to the FOPI and INDRA experimental data. It is found that, at $E_{\text{lab}}$ = 40 MeV/nucleon, the slope of the directed flow, the elliptic flow, as well as the nuclear stopping power ($vartl$ and $R_E$) can be well reproduced with calculations using $\mathcal{F}$=0.2, while $\mathcal{F}$=0.5 is required to reproduce these data at $E_{\text{lab}}$ =150 MeV/nucleon. These findings are consistent with the results deduced from the stopping data of protons by the INDRA collaboration. In addition, both the directed and elliptic flow parameters obtained with the FU3FP1 parametrization and $\mathcal{F}$=0.3 are quite close to each other, and sizable difference appears only at high transverse momenta. In general, calculations with the FU3FP1 parametrization fit the FOPI and INDRA data of both the collective flow and the nuclear stopping well, including their beam-energy dependence, while calculations with $\mathcal{F}$=0.2, 0.3, and 0.5 exhibit almost constant degree of stopping in Au+Au with beam energies increasing from 40 to 150 MeV$/$nucleon.

\begin{acknowledgments}
The authors acknowledge the support of the computing server C3S2 at the Huzhou University. This work is supported in part by the National Natural
Science Foundation of China (Nos. 11505057, 11375062, 11605270, 11675066, and 11747312), and the Zhejiang Provincial Natural Science Foundation of China (No. LY18A050002).
\end{acknowledgments}

\end{document}